\documentstyle[prl,epsfig,aps]{revtex}
\begin{document}
\twocolumn[\hsize\textwidth\columnwidth\hsize\csname@twocolumnfalse%
\endcsname
\title{Surface Kinetics And Generation Of Different Terms    
In A Conservative Growth Equation  }  
\author{S.V. Ghaisas} 
\address{
 Department of Electronic Science, University of
Pune, Pune 411007,
India}
 
\date{\today}
                  
\maketitle 
 
\begin{abstract} 
A method based on the kinetics of adatoms on a growing surface under 
epitaxial growth at low temperature in (1+1) dimensions is proposed to 
obtain a closed form of local growth equation. It can be generalized 
to any growth problem as long as diffusion of adatoms govern the 
surface morphology. The method can be easily extended to higher 
dimensions. The kinetic processes contributing to various terms in the 
growth equation (GE) are identified from the analysis of in-plane and 
downward hops. 
In particular, processes corresponding 
to  $h\rightarrow -h$ symmetry breaking term and curvature dependent 
term are discussed. Consequence of these terms on the stable to 
unstable transition in (1+1) dimension is analyzed. In (2+1) dimensions 
it is shown that an additional asymmetric term is generated due to the 
in-plane curvature associated with mound like structures. This 
term is independent of any diffusion barrier differences between 
in-plane and out of-plane migration. It is argued that terms 
generated in the presence of downward hops are the relevant terms 
in a GE.  
Growth 
equation in the closed form is obtained for various growth models 
introduced to capture most of the processes in experimental 
Molecular Beam Epitaxial Growth. Effect of dissociation is also considered 
and is seen to have stabilizing effect on growth. It is shown that 
for uphill current the GE approach fails to describe the growth 
since a given GE is not valid over the entire substrate.

\end{abstract} 
 
] 
\narrowtext 
\section {Introduction} 
Growth of solid phase from vapor is studied over many years due to its 
applications in various fields. In particular, epitaxial growth from vapor
 has given way to many technological advances through the development of 
solid state devices. Such a growth is known to be far from equilibrium
\cite{bar,kr1}. It therefore offers to study the non equilibrium(NE) 
phenomenon in a controlled fashion. Under these conditions, 
on the growing interface, processes 
such as island formation, dissociation, nucleation do not equilibrate, 
being limited by the insufficient mass transport. Resultant interface 
described by height function $h({\bf r},t)$, develops 
characteristic correlations on the surface. Study of space time evolution 
of these correlations constitute major aspect of understanding the NE 
behavior for such a growth. A continuum equation approach is used to 
understand this phenomenon\cite{bar,kr1}. Under most of the conditions 
used in growth from vapor, evaporation and the 
vacancy formation are negligible\cite{si}. A conservative growth 
equation satisfying Langevin equation  
 $\partial_{t}h+{\bf \nabla
 \cdot J}=F$ where 
 {\bf J} is current
due to adatom 
relaxation on the growing surface and $F$ is the 
average flux with 
 white noise, should  describe the time evolution of the 
height function. In order to understand the growth behaviour via 
 GE approach, it is necessary to establish the 
correspondence between surface kinetics and the terms appearing in 
the GE. This correspondence should corroborate with experimental observations.  
Experiments have produced results in a great variety
\cite{bar,kr1}. However adjusting experimental parameters to desired 
accuracy seems a daunting task so that effect of initial surface roughness, 
impurities in the flux and similar phenomena\cite{kr2} lead to results with 
less confidence to apply any particular GE, since there is a 
possibility of substantial contribution from these uncontrolled inputs. 
Under such conditions, computer simulations are expected to produce 
results that can be related to the physical processes used as inputs 
to the simulation giving more confidence in establishing the growth 
equation. In the present article, we have used computer simulations 
to verify the predictions of GE. The application of these 
results to real experiments is discussed in section (IV) under the 
presumption that the experimental conditions are same as assumed 
in the simulations. 
 
Based on the experimental observations and computer 
simulations\cite{len1,pv}, when relaxation is mainly by surface 
diffusion of adatoms,  minimum three terms are expected to 
contribute to the current: a slope dependent term, an asymmetric term 
and a curvature dependent term\cite{pv,po}. One of the early efforts 
\cite{vzw} used the Master Equation approach for the model involving 
desorption and column diffusion as the relaxation mechanisms for adatoms. 
This approach yielded slope dependent term due to desorption, and 
also the asymmetric term along with fourth ordered Mullin\cite{mul} type term.
The terms are obtained under small slope approximation so that the behavior 
at large slopes is not clear. Further, it is not possible to separate 
the kinetic processes, responsible for different terms. In the next 
section, we refer to some of the results from this work and suggest 
the contribution to GE with respect to in-plane and downward hops.  
Kinetic approach involving Arrhenius model or Burton, Cabrera and Frank 
(BCF)\cite{bcf} approach including nucleation and step edge barriers,   
could not yield a closed form for the growth equation\cite{kr1}. It 
could however relate the slope dependent term to diffusion of adatoms 
on the terrace of a step. To obtain asymmetric term additional assumption 
that density of adatoms should have quadratic slope dependence, was needed.
Both the terms are obtained under small slope approximation and behavior 
at large slopes is not clear. 
 The 
curvature dependent   
 term is interpreted as due to the step detachment\cite{len1} or due to the 
nucleation, based on dimensional arguments\cite{pv,po}. From the 
microscopic theory \cite{vzw}, surface diffusion will {\it always} 
create curvature dependent and asymmetric terms. Thus step detachment 
and nucleation, being result of the surface diffusion process, produce 
these terms. So far there has not been a clear correspondence 
established between various kinetic events and the terms that 
they are supposed to generate in a GE.  As has been noted in 
reference  \cite {po}, a systematic derivation of surface current 
is lacking.    
Thus, the correspondence between kinetic processes and terms in 
growth equations is not fully established in the case of NE growth. 
The work described in next few sections addresses this problem and 
said correspondence is established at least within the frame work 
of simple kinetic arguments.   

 In the present work we consider in-plane and downward hops  
as the basic relaxation mechanism that produce the necessary  
kinetic processes, giving rise to various terms mentioned above.
So far, role of such hops has not been explicitly considered in 
most of the work on growth. Distinctive effects of such hops have 
been considered in connection with RHEED specular spot oscillations
\cite{mg} and also while considering their effect on time exponent $\beta$
\cite{svgunpub}. In the method described in next section,    
 terms in the GE are obtained by considering the current, as affected by 
geometrical configuration relevant to the relaxation rules. Here we   
calculate the particle current in a heuristic way accounting for the 
processes generated by in-plane and downward hops by adatoms. This 
approach is less rigorous compared to previous ones\cite{kr1,vzw}. However it  
allows identify the terms in GE with kinetic processes 
directly. {\it This isolation of processes and identification of growth 
terms is the main contribution of the proposed method.}
  It is shown that 
{\it both} , slope dependent and asymmetric terms result from the {\it same} 
kinetic process. In (2+1) dimensions additional asymmetric term due to in-plane 
curvature is obtained. This term has relatively weak dependence on the SE barrier. 
Further, based on computer simulations in (1+1) dimensions 
 curvature dependent term is argued to be related to {\it downward} hopping.

 We apply this method 
to some of the well studied models namely DT model \cite{dt}, 
WV model \cite{wv}  mimicking the low temperature 
MBE growth and also for a model used by Lai- Das Sarma, LD model \cite{ld} to 
demonstrate the manifestation of Lai-Das Sarma- Villain (LDV) equation. For the 
DT model without noise reduction, closed form of the GE is not known, while 
 in the LD model, the GE is empirically related. 
Present 
method allows a closed form of GE for DT model for the first time 
and also shows how the rules in the LD model produce a term in growth 
equation that has similar behaviour as the Lai- Das Sarma- Villain equation.   
 
\section{Equation for Growth with Surface Diffusion}   
\begin{figure} 
\epsfxsize=\hsize \epsfysize = 1.0 in
\centerline{\epsfbox{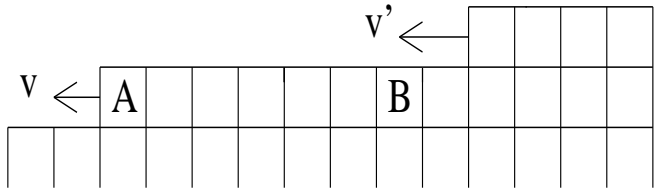}}
\caption{A typical step structure formed during 
growth along positive slope. $v$ and $v'$ are velocities of the steps.   
}     
\label{step}  
\end{figure}

 Consider growth on a 1 dimensional flat substrate
with lattice 
constant $a$ and steps developed after some initial growth. 
We will consider the situation depicted
in Fig. \ref{step}  
for obtaining various contributions to the current where steps 
are such that positive slope is obtained. Here the underlying assumption 
is that the rough or unstable surface will essentially consist of stepped 
regions mainly.   
Adatoms are randomly
 deposited on the substrate. Let $D_{s}$ be the
diffusion constant 
on the terrace, $l_{c}$ be the average distance that
an adatom travels 
on a terrace before encountering another adatom.
Detachment from steps 
or nuclei on the terrace is assumed negligible. Contribution of a hop to the 
current is considered {\it only when nearest neighbor ($nn$) configuration 
changes, during hop}. The configuration is not restricted to number of 
neighbors, but as required by the relaxation rules, it can be heights of 
neighbors, or local discrete slope or any such derivative. In the present 
context, detachment is not allowed, hence the number of {\it nn} matters. 
In fact we treat every individual process , so that when dissociation 
is allowed, its contribution is considered separately to the current while 
contribution to current in the absence of dissociation is retained.     
We further differentiate between the current due to 
downward and in-plane hops.
 Since the (1+1) dimensional surface essentially consists of such 
steps, kinetics of adatoms on such a surface will suffice to provide the 
essential description of growth in terms of a continuum GE for 
the given set of rules for relaxation of adatoms.   
Adatoms
hopping down the descending 
steps contribute to the downward current $j_{d}$ while
those hopping 
on the terrace can get attached to an ascending step
contribute to the 
in-plane current $j_{i}$. Referring to Fig. \ref{step} , the
adatoms reaching 
site $A$ and hopping down the step to the left
constitute $j_{d}$ and 
those reaching site $B$ and hopping to the right
toward the 
ascending step constitute $j_{i}$. The net current is
$j_{d}+j_{i}$. 
These are obtained in a mean field approach as follows,
 
    $j_{d(i)}$=( local density of site A(B))$\cdot$ (flux of
adatoms 
 approaching 
              A(B))$\cdot$(probability for hopping across
A(B)).
\begin{equation}
\end{equation}  

Since the relaxation of adatoms is through the 
diffusion on the terrace or across the step edges, 
the density of sites A and B is 
same as 
density of steps which is approximately given by
$\frac {|m|}{1+|m|}$, 
where $m$ is the local slope. In the limit $|m|\rightarrow 0$, this 
density approaches zero. However, Elkinani and Villain have shown 
\cite{ev} that a 'plane' substrate will actually consist of terraces 
of an average length say $l_{av}$. This will introduce an additional small
factor in the numerator of the expression for density. We will however 
consider the diffusion and deposition rates such that length 
$(D_{s}/F)^{1/4} < l_{av}$ \cite{kr1} so that after growth of few MLs terraces 
are shorter than $l_{av}$ and 
density of steps is $\frac {|m|}{1+|m|}$.       

In the absence of nucleation, lateral flux(LF) approaching site B or A is 
$\frac{\pm \hat n F}{2|m|a^{-1}}$, $a|m|^{-1}$ being the average local 
terrace width and  $\hat n$ is unit vector in the x direction.  
When terrace size is large enough, this flux is restricted due to 
the nucleation. The nucleation process will restrict the diffusion 
to  an average  length $l_{c}$ on a 
large terrace. As a result, even for very large terraces, the 
approaching flux LF is almost constant. The effect of nucleation 
is incorporated by introducing $l_{c}$ in the expression for 
LF as $\frac{\hat n F}{2(l_{c}^{-1}+|m|a^{-1})}$, so 
that for small slopes the expression is reduced to a constant value.
Let $P_{A}$ and $P_{B}$ represent the probabilities of hopping across 
sites A and B respectively. The Schwoebel length \cite{vill1,se} $l_{s}
\propto (P_{B}-P_{A})$ . In ref. \cite{ev} it is shown that, if there is 
large asymmetry between the sticking coefficients, distribution of 
diffusing adatoms on the terrace depends upon terrace width. This 
suggests that $P_{A}$ and $P_{B}$ may depend on $m$ for larger asymmetry
and shorter terraces. However when $P_{A}
\rightarrow 0$ and $P_{B}\rightarrow 1$ {\it i.e.} the case of large 
SE barrier, the nucleation becomes significant being proportional 
to the density \cite{bv}. As a result, the asymmetry of the density 
on the terrace is reduced, rendering $P_{A}$ and $P_{B}$ almost 
independent of the terrace width.  
The flux will depend only on the terrace width, so that  $P_{A}$ and $P_{B}$ 
will be independent of the width. Thus it suggests that the said 
dependence will be weak.  
We will however neglect this dependence since most of our discussion 
will be around $P_{A}\approx P_{B}$. Further, such an asymmetry is significant 
when $P_{B}>P_{A}$ since it will produce uphill current and large local 
slopes with shorter terraces. It will be shown that under these conditions, 
the continuum equation approach fails \cite{svg3}, so that these 
probabilities could be considered independent of $m$ where GE is valid.   
Hence from Eq.(1) the current is 
\begin{equation}  
   {\bf j}_{s}=\frac {\hat n |m|F(P_{B}-P_{A})
}{2(1+|m|)(l^{-1}_{c}+|m|a^{-1})}
\end{equation}

The LF approaching sites A and B is however 
 modified due to the relative motion of the steps. Consider the 
situation in Fig.\ref {step}  where $v$ is the velocity of the step bearing 
the terrace while $v'$ that of the higher one on the positive slope.
For $v'>v$ the terrace width is reduced, depleting the LF approaching 
sites A and B. The reduction in flux is $\propto \delta v$ where 
$\delta v=(v'-v)$. Adatoms hopping across upper step as well as those 
attaching in-plane, both contribute to the velocity of the step. Thus 
the velocity $v\propto j_{s}$ except that the coefficient $(P_{B}-P_{A})$
 is replaced by $(P_{A}+P_{B})$. Hence $\delta v\propto \frac{(P_{A}+P_{B})}
{l^{-1}_{c}+|m|a^{-1}}\partial_{x}\frac {|m|}{2(1+|m|)(l^{-1}_{c}+|m|a^{-1})}
$. Corresponding current {\it will not depend on} $(P_{B}-P_{A})$ since this 
part of the {\it flux is removed}
 from the LF. The current is therefore obtained 
by multiplying the LF by density of steps. In the process described , 
there is relative advancement of higher step.  With respect 
to the in-plane current , the step front advancement increases 
relative velocity. Thus effectively in-plane current will increase 
while downward current will decrease. It is this difference in currents 
represented by the current due to the lost flux.  
The relative movement of steps also causes increase in the local slope
 \cite {ercr}.  
This shows that the direction of the current due to the lost flux 
is same as uphill current which also 
tends to increase the local slope.  
 $P_{A}$ and $P_{B}$ are relative 
probabilities so that $P_{A}+P_{B}=1$. For $v'<v$ the terrace size is increased
. This situation is encountered for positive curvature. 
This produces a term assisting the LF. Using the same arguments as above, 
corresponding current has the same analytical form as the one for the case 
of $v'>v$ but with the coefficient $(P_{B}-P_{A})$ and in the direction 
opposite to it.    
Accounting for this effect the expression for 
the current becomes, 
\begin{eqnarray}  
  {\bf j}(x)&=&\frac {\hat n |m|F (P_{B}-P_{A})}{2(1+|m|)
(l^{-1}_{c}+|m|a^{-1})}
             \nonumber\\
&&+\frac {\hat n F(1-(P_{B}-P_{A}))}{4}\nonumber\\  
&&\partial_{x}
\left(\frac {|m|}{(1+|m|)(l^{-1}_{c}+|m|a^{-1})}\right)^{2}    
\end{eqnarray}
In the limit of small $m$, the current reduces to, 
\begin {eqnarray} 
 {\bf j}(x)&=&\hat n F(P_{B}-P_{A})|m|l_{c}/2\nonumber\\
&&+\hat n F(l^{2}_{c}/4)(1-(P_{B}-P_{A}))  
\partial_{x}(m^{2})
\end {eqnarray}  
The second term reduces for small slope conditions to ${\bf \nabla}(\nabla h)^{2}$. 
This term was derived using the BCF theory and 
assuming that at small slopes the particle density on the terraces 
depends on the even powers of local gradient \cite{kr1,pv}.  
 Further, in ref. \cite{ld} 
it was conjectured that such a term can arise due to the differences in 
the velocities of the steps near the top and the bottom of a profile.  
 In the limit of large slope, $|m|a^{-1}\gg l^{-1}_{c}$  
the second term is proportional to $\frac {1}{m^{3}}\partial_{x}m$, 
which was derived in the large slope limit in reference \cite{pv}. 
Thus the geometrical dependence of the symmetry breaking term 
in Eq.(3) exactly matches with previously derived two terms in the 
small and large slope limits. This shows that the asymmetric term in 
equation (3) appropriately interpolates through the limits of small and 
large slopes implying the correct analytical form of the term. 
Further, 
under the infinite SE barrier, {\it i.e.} $P_{B}=1$ and $P_{A}=0$, second 
term in above equation becomes zero. From computer 
simulations in (1+1)-d and in (2+1)-d it is seen that for the former case   
\cite{kr3,cak} , growth morphology is symmetric 
with respect to $h\rightarrow-h$ transformation while for the later case 
it is asymmetric. It has been suggested \cite{po1} that asymmetric term 
does not go to zero, it however decreases faster than slope dependent 
term which dominates to show symmetric profile in (1+1) dimensions under 
infinite SE barrier. Our analysis however shows that, in (1+1) dimensions, 
the asymmetric term is absent. In (2+1) dimensions, an additional effect 
of in-plane curvature gradient generates the asymmetric term.  
Under 
most of the circumstances, one obtains finite structures on the surface. 
In particular 
when mound like structures are formed, steps on the mounds have  
{\it in-plane curvature}. The in-plane curvature 
is given as $\kappa(x,y)=\frac{ h_{xx}h_{y}^{2}-2h_{xy}h_{x}h_{y}+h_{yy}
h_{x}^{2}}{(h_{x}^{2}+h_{y}^{2})^{3/2}}$, 
where, $h_{x}, h_{y}$ are derivatives of the height function $h(x,y)$ 
w.r.t. $x$ and $y$ respectively. 
Consider a region where steps are forming concentric 
arcs ,  $P_{B}\ge P_{A}$ and surface diffusion is 
isotropic . Under such conditions, the 
inward flux is proportional to $R^{-1}$, where 
$R^{-1}=|\kappa|$ is 
the radius of curvature at the 
point under consideration. In such region, the velocity along $\hat n$
is along the radius of curvature. In the presence of 
 $R^{-1}$ dependence, velocity 
gradient is created due to the gradient in curvature. 
Thus, by applying previous considerations for obtaining the current, 
we find that current due to the difference in in-plane curvature across 
consecutive planes is 
\begin{equation} 
j_{cur}=\frac{\nabla h Fa(\frac {\nabla h}{|\nabla h|}\cdot \nabla)|\kappa
(x,y)|}{2(1+|\nabla h|)(l_{c}^{-1}+a^{-1}|\nabla h|)} 
\end{equation} 
This term will be present {\it in addition} to the one due to the curvature in 
the height profile. Further, since for mound like structures, in-plane 
curvature gradient is always positive implying that flux is removed. 
Hence , in the expression for $j_{cur}$, $P_{A}$ or $P_{B}$ do not appear 
explicitly. For small slope conditions, term in growth equation corresponding 
to the $j_{cur}$ is $\nabla \cdot(\nabla h Fa(\frac {\nabla h}
{|\nabla h|} \cdot 
\nabla)|\kappa|)$. Under scaling transformations, this term gives $z-4$, 
which is same as $\nabla^{4}h$ term, but differs in the $h\rightarrow -h$ 
symmetry. This shows that for the growth near tilt independent current in 
(2+1) dimensions, mounds grow with $t^{1/4}$ while $h\rightarrow-h$ 
symmetry is broken. Thus, our analysis shows that, for an infinite 
barrier, in (1+1) dimensions, asymmetric term is zero while in (2+1) dimensions 
in-plane curvature gradient generates asymmetric term. In fact, in almost 
all (2+1) dimensional growths, asymmetry due to this term is unavoidable 
if mound formation occurs.

We further argue that a curvature dependent current must be present 
in any adatom relaxation process that involves downward hops across 
the descending step edges. This argument is based on the observation 
that, in a (1+1) dimensional simulation, 
if adatoms are restricted completely to 
the in-plane hops (infinite SE barrier) then correlations do not 
grow beyond the diffusion length. This results in to the 'wedding cake' type 
 morphology with fixed size of the 'cakes' that do not grow in time 
\cite{kr3,cak} as discussed above.  
 On the other hand, when such hops
 are allowed, correlation length for stable growth and mound size for 
unstable growth increases in time\cite{svgunpub}. This observation allows 
one to conclude that height-height (h-h) correlations {\it increase only 
in the presence of downward hops}. The microscopic theory \cite{vzw} 
predicts fourth ordered and asymmetric term to be present irrespective 
of in-plane or out of-plane hops. Thus, accordingly downward hops  
produce these terms, however in plane hops also generate 
these terms. In order to differentiate between the role 
of in-plane and downward 
hops in generating these terms, consider a rough surface, obtained 
after deposition of several monolayers. 
We consider an in-plane hop and a downward hop. Let $W_{i}$ and $W_{f}$ be the 
initial and final widths with  $W=(1/N)\sum_{j} (h_{j}-\bar h)^{2}$. 
Also let $G_{2i}(|l-k|)$ and $G_{2f}(|l-k|)$ be initial and final height-height 
correlations. Consider a hop from site i to i+1. The in-plane hop gives 
$W_{i}=W_{f}$ , while downward hop gives $W_{i}-W_{f}=4a^{2}$. Thus, the 
width is reduced due to the downward hop. In order to see the effect on 
h-h correlations, we consider only contributions from participating sites. 
Thus, contribution to $G_{2i}(|i-j|)$ will be from sites at $h_{i}$,
$h_{i+1}$, $h_{j}$ , $h_{j+1}$ and corresponding reflection sites 
in i and i+1. The difference $G_{2i}(|i-j|)-G_{2f}(|i-j|)$ 
due to one set of sites is $2a^{2}-2a(h_{j}-h_{j+1})$. The ensemble 
average for this process will yield $2a^{2}$ as the difference. Same 
contribution will appear from reflection sites. Thus, $G_{2}(|i-j|)$ 
is reduced by $4a^{2}$ by a downward hop. On the other hand, in-plane hop 
does not change $G_{2}(|i-j|)$, as can be verified by applying same 
procedure. This shows that, {\it in order that correlations grow in time, 
downward hops are necessary.} The in-plane hops reduce the deposition 
noise in a plane. Consider a large flat surface with very small coverage. 
Every event of an adatom getting captured by a nucleus, or another adatom 
causes decrease in $G_{2}(1)$. For an island of $m$ atoms on a 1- dimensional substrate, 
$G_{2}(1), G_{2}(2),...G_{2}(m)$ decrease from its value for $m$ isolated atoms.
 This indicates reduction of noise in a plane. 
In fact, this process develop correlations over the diffusion 
length $l_{d}$. In ref. \cite{vzw}, fourth ordered and asymmetric terms 
are generated whenever near neighbor configuration is changed in a hop. 
This is consistent with the reduction in $G_{2}(m), m\le l_{d}$
 with hops leading to 
increase in near neighbors. Thus in-plane hops do generate fourth order 
and asymmetric terms, however, these exclusively operate within 
$l_{d}$, reducing deposition noise in a plane. The processes, 
such as nucleation and step attachment or detachment \cite{vill1,po} 
are suggested to generate fourth ordered term. Above discussion leads 
to the conclusion that such a term would operate only within the 
plane and not across different planes. This shows that {\it in a growth 
equation, terms generated by the kinetic processes involving downward 
hops are the relevant terms.}This also suggest that upward hops will 
{\it decrease} the correlations. We do not consider in the present work 
associated effects, as we restrict to analyzing low temperature growth.    
Note that adatoms crossing site $A$ are 
hopping downward and lead to the EW type term. Hence if the downward 
hops are allowed but the current is tilt-independent, then it may be 
expressed as a linear combination  $a_{1}\nabla^{3} h+a_{2}\nabla^{5} h
+...$ including the nonlinear terms of the form ${\bf \nabla}(\nabla^{3} h)
^{2}$. Here, $a_{1}, a_{2},...$ are such that the growth is stabilized.  
We will retain only $\nabla^{3} h$ in the current corresponding 
to our minimal GE. Thus the form of the current corresponding 
to the minimal GE in (1+1) dimensions is 
\begin {eqnarray}
  {\bf j}(x)&=&\frac {\hat n |m|F(P_{B}-P_{A})}{2(1+|m|)(l^{-1}_{c}+
|m|a^{-1})} 
+\frac {\hat n|m|F(1-P_{B}+P_{A})}{4}\nonumber\\    
&& \partial_{x}\left(\frac {|m|}{(1+|m|)(l^{-1}_{c}+|m|a^{-1})}\right)^{2}
+\nu \frac{\partial^{3}h}{\partial x^{3}}  
\end {eqnarray} 

The first term has been studied widely as the stable growth mode
\cite{ew,fam} and as the unstable growth mode\cite{len}. We aim to study 
tilt independent current models here. This will allow exclusive effect 
of asymmetric term to be studied.     
With this view we have performed simulations of a (1+1) dimensional 
solid-on-solid model with {\it no diffusion bias}. This will produce growth  
with tilt independent (TI) current. 
A fourth ordered 
equation was earlier proposed by Villain \cite{vill1} for similar situation.  
Under this condition 
$P_{A}=P_{B}$ and first term vanishes in the Eq(3). The resultant 
GE in the moving frame with growth front is of the form, 
\begin {equation}
\partial_{t}h=-\nu\frac {\partial^{4}h}{\partial x^{4}}+\nu_{a}\partial_{x}
^{2}\left
(\frac {m}{(1+|m|)((l^{-1}_{c}+
|m|a^{-1})}\right)^{2}+\eta 
\end {equation} 
where, $\nu_{a}$ is the appropriate constant for the asymmetry term 
and $\eta$ is white noise associated with deposition flux  
with $<\eta(x',t')\eta(x,t)>=D\delta(x'-x)
\delta(t'-t)$. In the limit of small slopes, 
renormalization group (RG) analysis 
shows that the roughness exponent $\alpha=1$ and the roughness  
evolves  with the exponent $\beta=1/3$\cite{ld}. For large slopes, 
 when terrace width is very small (of the order $ a$), the multiplying 
length in the expression $l\frac{\partial j(x)}{\partial x}$ is no more 
represented by $(l_{c}^{-1}+a^{-1}|m|)$ but is to be replaced by $a$. This 
leads to asymmetric term with $z=4$ that describes DT model \cite {dt} in the 
absence of noise reduction. We discuss the GE for DT model in section III A.
  Here we note that, asymptotically Eq. 7 will reduce to a GE describing 
DT model, hence exponent $\beta$ should cross over from a value of 
$1/3$ to $3/8$. 
 In the next section we describe a solid-on -
solid growth model that mimics the relaxation by surface diffusion. The 
relaxation rules are consistent with the processes giving rise to different 
terms in Eq. (6). These results will help establish the relationship between 
process $\rightarrow$ 
term in GE . In section V, we apply this method to predict 
GE for other models.   

Extension of this equation to (2+1) dimensions is possible by 
similar kinetic considerations. For isotropic diffusion, same 
form as Eq.(6) is obtained with $\hat n={\bf \nabla}h/|\nabla h|$ 
and replacing length-derivative product in 
asymmetric term by $\frac{\nabla h\cdot \nabla}{(l_{c}^{-1}+
|\nabla h|)|\nabla h|}$ and adding the asymmetry term due to the 
in-plane curvature gradient. 
 For small slopes 
one obtains $\nabla(\nabla h)^{2}$ term\cite{ld}along with 
$\nabla \cdot(\nabla h (\nabla h \cdot \nabla)|\kappa|)$,  which seems 
to describe many experimentally observed growth roughness measurements 
from vapor\cite{krim}. In particular, large number of experiments 
show the roughness exponent $\alpha$ between 0.65-1.0 predicted by 
these two terms. 

\section {Growth Model}    
\subsection {Growth with surface diffusion and dissociation}  
Corresponding model is as follows.
Atoms are rained on a 1-d substrate of length $L$ 
randomly with constant flux. On deposition a given adatom is allowed to 
hop n times, as in a random walk. The hops can be biased through a parameter 
$p$. Thus, $p=1.0$ is the growth with infinite positive
SE barrier, while $p=0.0$ is with infinite negative barrier.  
We have however 
kept $p=1/2$, {\it i.e.} no bias
condition for most of the cases.  
If the hopping adatom acquires two or more $nn$,  
 before n hops are exhausted, the adatom stays there 
permanently. If n hops are exhausted without any encounter, it stays 
permanently at the last position occupied after n hops. If a single 
$nn$ is acquired before $n$ hops are exhausted, another parameter 
$q$ is called. $q$ decides fraction of such events, where adatom 
will dissociate from its neighbor. For $q=0$, detachment is completely 
suppressed. Under this condition detailed balance is not obeyed.   
As usual, $q$ is compared with the random 
number to decide whether detachment can take place or not.  
We have extended the same model in (2+1) dimensions as well. Besides the 
parameters $p$ and $q$ controlling the hops across the step edge and 
away from the edge respectively, additionally edge diffusion has been 
included \cite{tl}. It is considered to be intra planer process. We also employ 
the noise reduction method\cite{wl} where ever needed. 
In this method, after deposition 
an adatom is allowed to make hops as per the rules until it finds 
the location for the incorporation. However, instead of actually 
incorporating the atom at that position, a counter at that position 
is increased by unity. A given position is filled only when the counter 
exceeds certain pre decided number. The method has been successful in 
bringing out the correct nature of the growth at earlier times in simulation
\cite{dtpin}. We find that when number of allowed hops are large enough, 
noise reduction occurs in the diffusion process during initial growth.  
 
In (2+1) dimensions , it may be noted that the slope 
independent current {\it is not 
obtained} for $P_{A}=P_{B}$ due to the different number of configurations  
 for in-plane and downward hops \cite{dtpin}. Thus for the same 
set of rules, results in (1+1) dimensions will differ from those in 
(2+1) dimensions as noted in ref.\cite{dtpin}. 

The present model includes 
the physical processes dependent on the surface diffusion. It however 
differs from kinetic Monte Carlo (KMC) method, usually employed in such 
simulations. Firstly the detailed balance is accounted in KMC since 
dissociation from the steps or nuclei is allowed as per the activation 
barrier for that event. The dissociation from steps is opposite to 
attachment from site $B$ in Figure 1. Hence, these events constitute 
additional downhill current. This current is however {\it independent 
of terrace width} and depends only on the density of edges
(the details are discussed in sec. IV B). As a result 
, the slope dependent first term in equation \label{} decreases with 
slope while dissociation dependent term doesn't. This leads to tilt 
independent current. We will illustrate this effect in the next section.    
KMC also 
allows accordingly, upward hops and edge diffusion. Thus based on the 
considerations of contributing currents, KMC method would tend to 
TI current growth rather than a true uphill current.   
The present model is however
computationally convenient and allows variation of parameters in such a 
way that isolation of processes and their effect on GE can be 
studied. By adjusting parameters in our model, downhill, zero or 
uphill current can be maintained during growth simulation.  
For comparing the predictions of GE with simulations, 
we have measured width $w_{2}$, 
height-height correlations $G({\bf r},t)$ and skewness $\sigma$ 
 where, $w_{2}=\frac{1}{N}
\sum_{i}(h_{i}-\bar h) \sim t^{2\beta}$ and $G({\bf
r},t)=\frac{1}{N} 
\sum_{{\bf r'}}(h({\bf r}+{\bf r'},t)-h({\bf
r'},t))^{2}$.   
 The skewness
$\sigma=w_{3}/w_{2}^{3/2}$ ,
where $w_{3}=\frac{1}{N}\sum_{i}(h_{i}-\bar
h)^{3}$\cite{kim}.

\section {Results} 
\subsection{TI Current Without Dissociation} 
\begin{figure} 
\epsfxsize=\hsize \epsfysize= 3.0 in
\centerline{\epsfbox{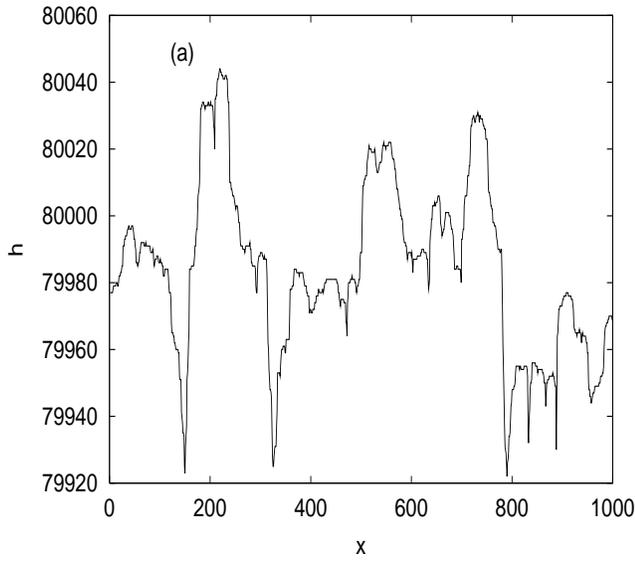}}
\epsfxsize=\hsize \epsfysize = 3.0 in
\centerline{\epsfbox{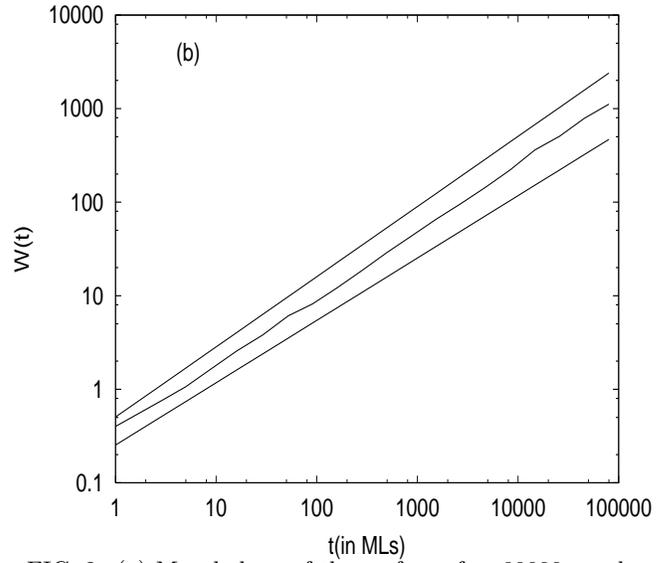}}
\caption{(a) Morphology of the surface after 80000 
number of layers.  
 (b)Plot of width as a function of time. Straight 
lines with slope $3/4$ and $2/3$ are drawn for reference. 
The substrate size is 10000 and SE barrier parameter $p$ is 0.5
({\it i.e.} no SE barrier).  
}    
\label{morfw}  
\end{figure}

 From the derivation 
of the expression (6), it is clear that TI current is 
obtained when $p=0.5$ and $q=0$, allowing $P_{A}=P_{B}$. Thus, mainly two terms 
are expected to contribute in GE, the $\nabla^{4}h$ and the asymmetric term.    
Presence of $\nabla^{4}h$ term is verified from the flatness of the 
saturated width for small $L$. We have chosen, n=10 giving $l_{c}\approx 3$. 
The saturated width is flat almost up to 5$l_{c}$, showing that $\nabla^{4}h$ 
dominates at small lengths\cite{pv}. 
 Fig. \ref{morfw}(a) shows the morphology of the interface after 80000MLs 
are grown. As predicted by the Eq.(7), the asymmetry is 
evident in the figure with $\sigma=-0.31\pm0.05$.  
Fig. \ref {morfw}(b) shows plot of $w_{2}$ in time. 
We obtain initially $\beta$ around 0.33 that attains a   
value of $0.35\pm0.015$. Initial value of 0.33 is due to 
the small slope region of Eq. 7 
that predicts a value of 1/3 (compare data in the  
region from 10 ML to 200 ML in the figure with the line having 
slope of $2/3$). 
Correspondingly h-h correlations lead to the roughness 
exponent   
$\alpha$ that increases from 0.5 to $0.75\pm0.01$ 
over a growth of $10^{3}$ to $4.10^{6}$ MLs. Clearly, the $\alpha$ tends 
to unity asymptotically on large substrates. The value of $\alpha$ 
from saturated width and for small $n$ is $1.35 \pm 0.1$.  
These results indicate 
that most of the morphological features of the growth with diffusion without 
detachment are captured by the current expression in Eq.(6). As is mentioned 
in section II, a slow cross over from $\beta=1/3$ to $\beta=3/8$ is observed. 
The exponent $\alpha$ from $W_{sat}$ is also close to the predicted value 
of 1.5 \cite {dt}. Thus the model in section III 
represents the GE given by Eq. 7 
confirming the association of kinetic processes with the terms in GE. 
It also shows that diffusion of the 
adatoms roughens the growing surface. Diffusion bias causes additional 
effects in terms of stability or instability of growth. 
In particular, if the bias is varied from extreme -ve SE barrier 
to extreme +ve SE barrier, a stable$\rightarrow$ unstable transition 
is observed. In this transition however $h\rightarrow -h$ symmetry is
 broken asymptotically. Note that for -ve SE barrier $\nu_{2}\nabla
^{2}h$ term dominates with +ve value of $\nu_{2}$\cite{ew}, so that 
asymptotically, asymmetric term becomes irrelevant rendering  
$\sigma=0$. At exactly zero 
SE barrier, finite value of $\sigma $
 is obtained. $\sigma$ can be  
regarded as the symmetry parameter, that changes abruptly 
 at the transition point.  
 Thus the growth transition is like 2nd order phase transition. This 
transition is however , a result of not complying with detailed balance. 

\begin{figure} 
\epsfxsize=\hsize \epsfysize= 3.0 in
\centerline{\epsfbox{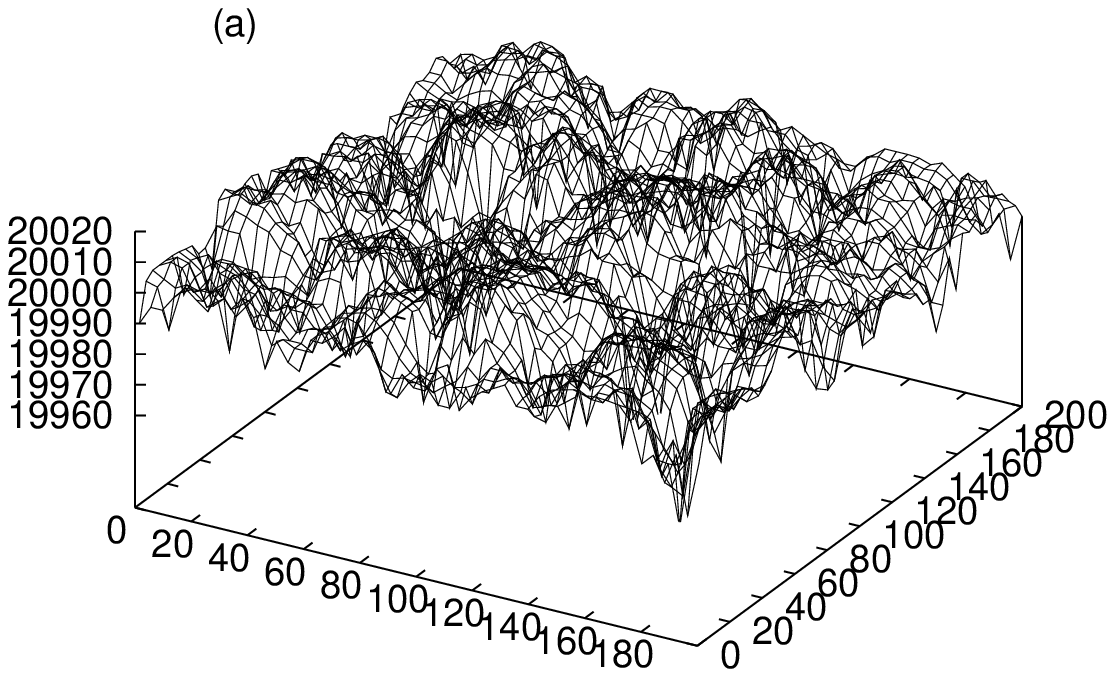}}
\epsfxsize=\hsize \epsfysize = 3.0 in
\centerline{\epsfbox{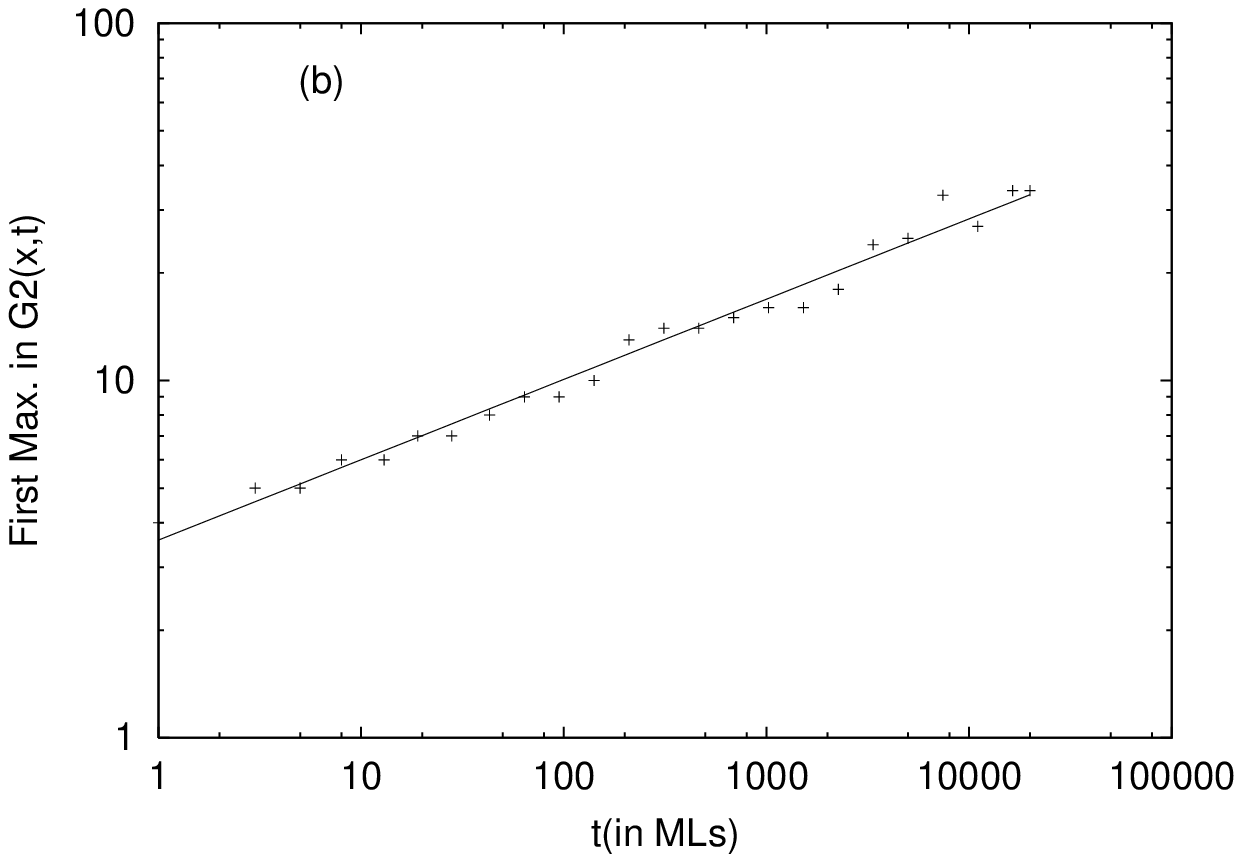}}
\caption{
  The growth model with small +ve SE barrier but without edge  
 diffusion. The growth is over 200X200 substrate 
size with SE barrier parameter $p=0.54$ showing small +ve barrier 
to compensate for the larger number of configurations, available 
for downward hops.   
(a) Morphology of the surface after 2000 
number of layers.  
 (b)Plot of measure of mound size as a function of time. The slope 
obtained is 0.23$\pm 0.02$. 
}
\label{ti2d}     
\end{figure}

As mentioned earlier, in (2+1) dimensions, TI current growth could not be 
obtained by putting $P_{A}=P_{B}$. We could attain a situation close 
to TI current growth by setting the parameter $p$ to a value 0.54,  
and without edge diffusion. Results in Fig\ref{ti2d}(a) 
show the morphology, where mound like structures are evident while 
Fig\ref{ti2d}(b) shows plot of position of first maximum in $G({\bf x},t)$ Vs. 
time. The $\beta=0.26\pm0.03$ while slope of the curve 
representing exponent $1/z$, in Fig\ref{ti2d}(b) is 
$0.23\pm0.02$ and  $\sigma=-1.12\pm0.1$.  
Clearly, an indication of dominant fourth ordered term with asymmetry.  
As has been discussed previously, it is the term generated from the 
the current $j_{cur}$ in Eq. 5, governing the growth dynamics.

\subsection{Effect of Dissociation} 

\begin{figure} 
\epsfxsize=\hsize \epsfysize= 3.0 in
\centerline{\epsfbox{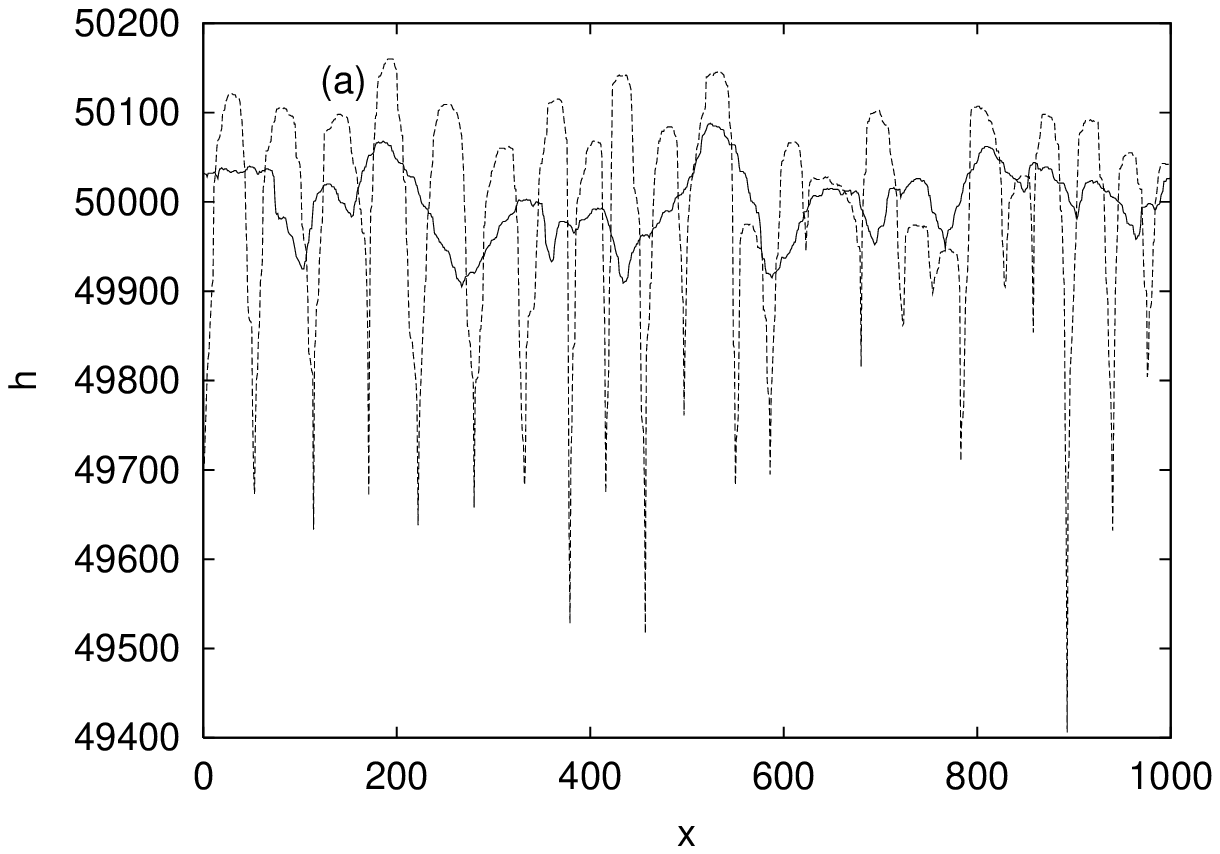}}
\epsfxsize=\hsize \epsfysize = 3.0 in
\centerline{\epsfbox{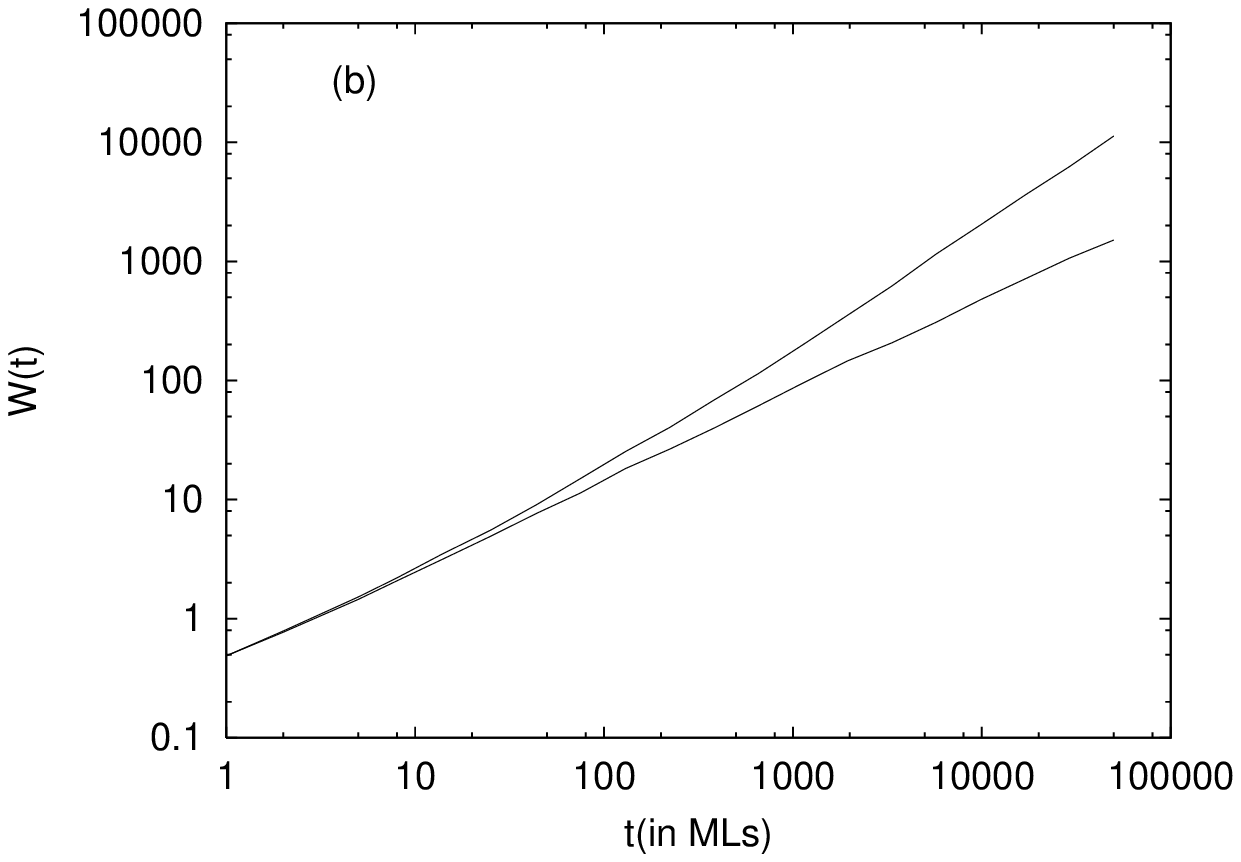}}
\caption{Comparison of growth model results, with dissociation
 and without, in (1+1) dimensions. Parameter $q$ decides the fraction 
of adatoms with single neighbor dissociated if encountered during hopping.  
(a) Morphology of the surface after 50000 
number of layers. Dotted curve represents morphology in the absence 
of dissociation while solid one is in the presence of it.   
 (b)Plot of width as a function of time. The $\beta$ value with dissociation 
is $0.377\pm0.007$, while that without dissociation increases to 1/2.   
}
\label{morphq}     
\end{figure}

\begin{figure} 
\epsfxsize=\hsize \epsfysize= 3.0 in
\centerline{\epsfbox{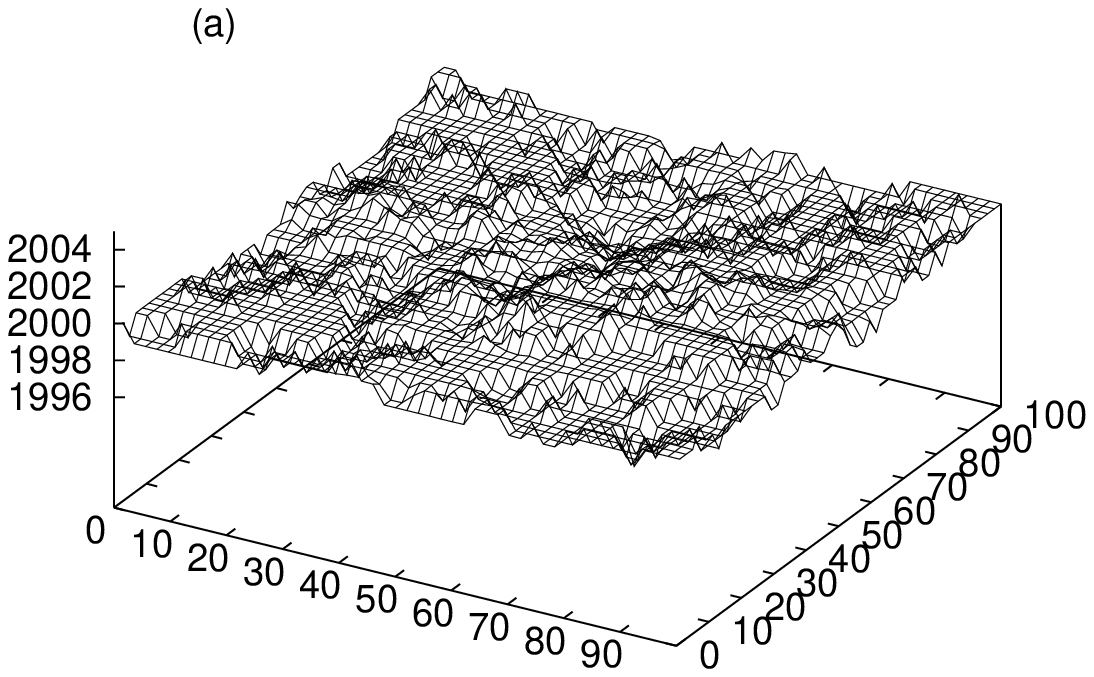}}
\epsfxsize=\hsize \epsfysize = 3.0 in
\centerline{\epsfbox{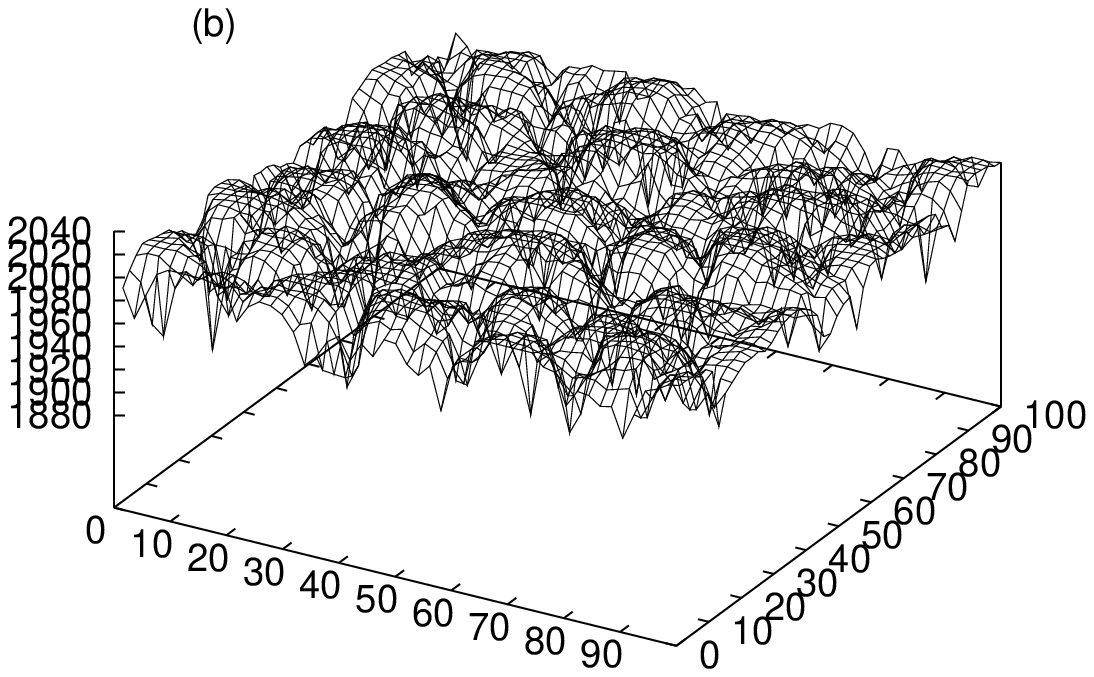}}
\caption{Comparison of growth model results, with dissociation
 and without in (2+1) dimensions. Parameter $q$ decides the fraction 
of adatoms with single neighbor dissociated if encountered during hopping.  
(a) Morphology of the surface after 2000 
number of layers with dissociation ($q=0.4$). Note the absence of mounds
in this case.   
(b) Morphology of the surface after 2000 
number of layers without dissociation ($q=0.0$). Note the  mounds
in this case.   
}
\label{morph2q}     
\end{figure}

As we have mentioned earlier, dissociation from the steps introduces 
a term of the form $-\hat n \frac {c'F|m|}{1+|m|}$ in the current, as a 
downhill contribution. $c'$ is fraction of the incident flux contributing 
to dissociation.  
This current does not decrease with $m$. Hence, 
in the presence of SE barrier and/or edge diffusion, with increasing 
average local slopes, zero tilt current is attained and analysis of 
previous section applies. To illustrate this point, in our (1+1) dimensional 
model, we introduce, SE barrier by assigning $p=0.7$. In order to simulate 
the dissociation effect, we take $q=0.01$, {\it i.e.} one hundredth of the 
total encounters with adatoms having 
single in-plane neighbors are allowed to hop in-plane 
or downward across the step edge. The results are displayed in 
Fig.\ref{morphq} a) and b) showing morphology and roughness evolution 
respectively.    
For the sake of comparison, curve corresponding to $q=0.0$ is also included. 
As seen from the figure 4 b), the $\beta$ value for $q=0.01$ is same as the one 
for zero tilt current case within statistical error. The $\beta$ for $q=0.0$ 
increases to 1/2 showing the instability. The argument is true for higher 
dimension as well. However, present model is not designed to account the 
effect of detailed balance. Thus, TI current is obtained for a 
certain set of parameters only and does not evolve as in the case of 
KMC simulation accounting for all the processes relevant to 
the detailed balance.   
In (1+1) dimensions, $(0)nn\leftrightarrow(1)nn$ and
 $(1)nn\leftrightarrow(1)nn$ are  
main processes during surface equilibration. Hence arriving 
at TI current is possible in (1+1) dimensions with our 
model that includes these processes controlled through parameters $p$ and $q$. 
 In (2+1) 
dimensions, the attachment-detachment  processes are many due 
to the different possible configurations. Present model does not allow all such 
processes. Thus, tilt independence becomes difficult to attain through 
variation of model parameters. 
 Hence in (2+1) dimensions we illustrate the dissociation 
effect mainly in the form of stable logarithmic growth.   
Fig.\ref{morph2q}(a) and (b) shows morphology for 
the case with and without dissociation in (2+1) dimensions respectively. 
The effect of 
dissociation is seen as the logarithmic growth as against a mounding with 
$\beta=1/2$ in the absence of it. The behavior of (2+1) dimensional model 
under TI current conditions can be predicted from the form of growth 
equation. 
 If the steps are straight, then 
LDV type  term will dominate giving  
, $\alpha=2/3$, 
and $\beta=1/5$ \cite{ld}. 
However, if mound like structures are formed , asymmetric term due to 
the in-plane curvature gradient (Eq. 5) will be operative leading to 
$z=4$ and $\beta=1/4$. 
 Siegert and Plischke \cite{sp} have considered a symmetric term 
to explain the pyramid like structures giving $z=4$ and $\beta=1/4$.

\section{Growth Equations for Other Models} 
\subsection{DT Model}  
\begin{figure} 
\epsfxsize=\hsize \epsfysize= 3.0 in
\centerline{\epsfbox{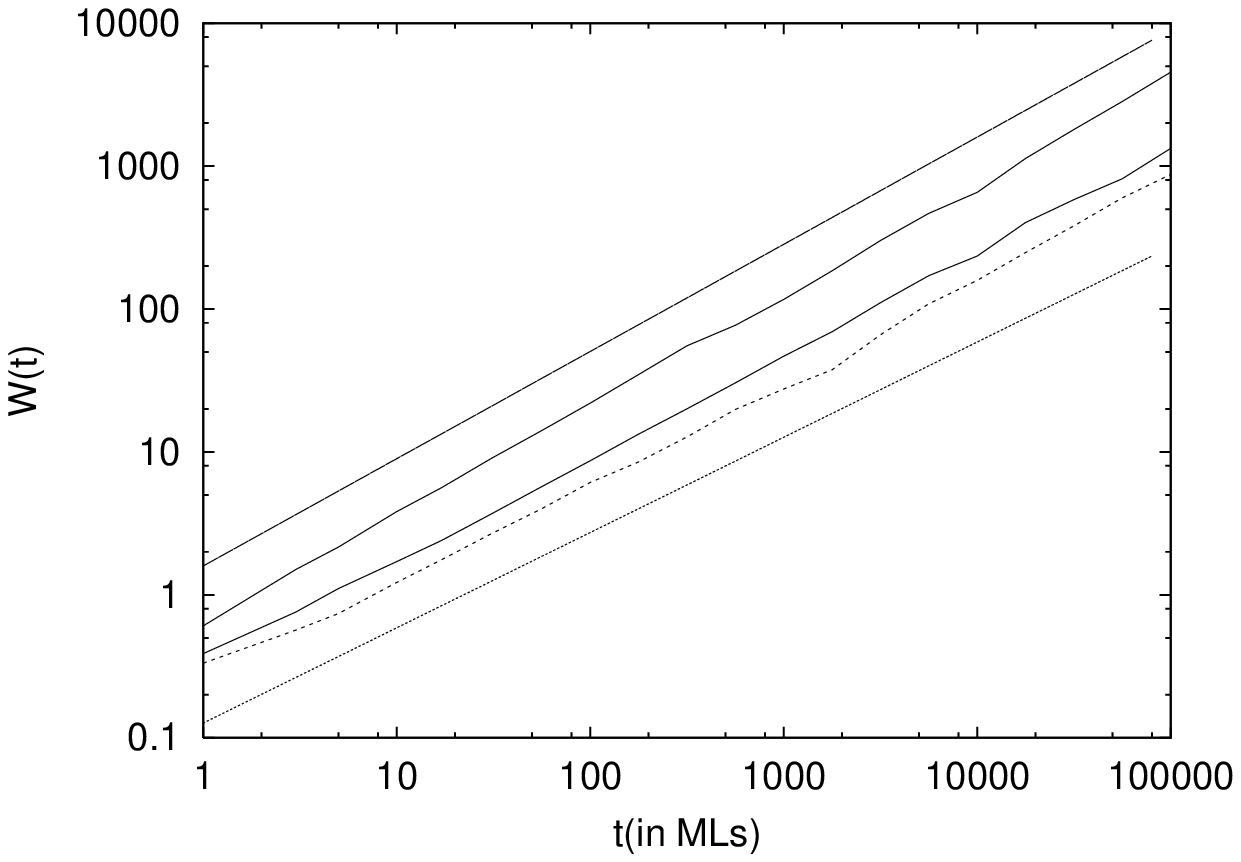}}
\caption{Plot  
  of width as a function of time for the (1+1) dimensional 
model described in section III for different number of hops $n$. Straight 
lines with slope $3/4$ and $2/3$ are drawn for reference as top curve 
and bottom curve respectively. In between, the curves from top correspond 
to $n=1,10$ and 25 number of maximum hops.  
The substrate size is 10000 and SE barrier parameter is 0.5 {\it i.e.}
no SE barrier.   
}    
\label{nbeta}  
\end{figure}

As mentioned earlier, the present method for obtaining current from 
kinetic considerations appropriately brings out the geometrical 
dependence in GE. We have applied this method to one 
of the stochastic growth models proposed to capture the essential 
features of low temperature molecular beam epitaxy (MBE) the 
DT model\cite{dt}. Based on noise reduction technique, the simulations 
of this model\cite{dtpin} confirm that,   
{\it 1)} exponent $\beta=3/8$ with noise reduction factor unity , while 
$\beta=1/3$ with reduction factor 10, {\it 2)} the morphology is asymmetric 
with $\sigma \approx -0.5$ , {\it 3)} the current is tilt independent
and {\it 4)} $\alpha=1.4$ and 1.0 with noise reduction factor unity 
and 10 respectively\cite{kr1,dtpin}.  
The relaxation rules for 
adatom in this model allow it to hop only when it is deposited 
at site A or B (see Fig. \ref{step}). 
Also only downward hop is allowed, if deposited at A 
and toward the step, if deposited at B. If it has choice of sites 
A,A or B,B or A,B etc. on two neighboring sites, then it will hop randomly 
to the 
left or right. Applying the considerations for obtaining current 
for this situation shows that the LF  
approaching sites A or B in the present set of rules is $\frac {F}
{a^{-1}+|m|a^{-1}}$. For this model, $l_{c}=a$ since only single 
hop in definite direction is permitted.  
Further, the LF is affected by relative motion of steps only when 
sites A and B differ by a lattice constant. In Eq(3), the velocity 
gradient is considered over a length of $(l^{-1}_{c}+a^{-1}|m|)^{-1}$ 
which is terrace width.
For the DT model, this length is $a$, the lattice constant, since the 
relative motion of steps for above set of rules can affect the LF only 
for this short terrace or lower. However lower terrace sizes  are not possible,
introducing effect of 
discretization as length cannot be reduced further. If n=1 in our model,
described in section III,  
the situation is not discriminated by the present analysis. We expect 
that exactly same equation governing DT model should be applicable. 
In fact, when local slopes increase in time, the effect of discretization 
shows up and exponents pertinent to DT model characterize 
the model asymptotically as seen in section IV A. Accounting for 
this effect, the expression 
for the current is then, 
\begin{eqnarray} 
J_{DT}(x)&=&\nu \frac{\partial^{2}m}{\partial{x^2}}+ \frac {\hat n
Fa^{2}}{2}\nonumber\\
&&\frac {m}{1+|m|} \partial_{x}\frac {m}{(1+|m|)^{2}}
\end {eqnarray} 
 From the rules it is clear that $P_{A}=P_{B}$. The GE 
corresponding to this current in the moving frame will be, 
\begin{equation}
\partial_{t}h=-\nu\frac{\partial^{4}h}{\partial{x^4}}
+\nu_{a} ' \partial_{x} \frac {m}{1+|m|} \partial_{x}  
\frac {m}{(1+|m|)^{2}}+\eta
\end {equation}
where, $\nu_{a} '$ accounts for various constants in the corresponding 
expression for current. 
The power counting in this equation
 leads to , $z=4$ from the first term, and $z=1+2\alpha$ for 
large slopes  corresponding to
the second term which is expected to be operative 
mainly over large local inclinations. The relation obtained from 
 the second term is exactly same as 
the one obtainable from the noise term $\eta$. For $z=4$ all the 
terms are marginal. This implies that  
 $z=4$ and $\beta=3/8$. The second term is symmetry breaking. 
Hence, above equation accounts for all the observed facts mentioned 
above in the 
simulation of DT model. Since the GE for DT model is obtained in the 
discretization limit of surface diffusion model, the asymptotic limit 
of surface diffusion model without dissociation is DT model.  
 Fig.\ref{nbeta} shows plot of $w_{2}$ Vs time for 
our (1+1) dimensional model for different values of n. As is expected, 
for n=1, $\beta$ is 3/8 while as n increases, it crosses over to this 
value at later time. Thus these results clearly demonstrate the effect 
of discretization in growth. 

With large noise reduction factor however, the observed behaviour 
of this model corresponds to LDV type GE \cite{dtpin}. We have 
seen that the form of GE with large enough terraces is indeed LDV type, 
as in Eq.7. 
Effect of noise reduction technique is to reduce the nucleation noise. 
In the process, longer terraces are created and {\it maintained} during 
growth. Thus, discretization limit is never reached leading to LDV type 
behaviour. In both cases, the current is TI, so the universality 
of this model is 'zero current universality'. It is a degenerate 
case since $z=3$ and $z=4$ are both possible for the same model.      

In (2+1)dimensions, with above rules for adatom 
relaxation, the local density of sites A and B need not be equal  
 since, fluctuations in step edges render configurations that 
show bias for sites A or B. As a result, slope dependent current 
will dominate the growth changing the universality class with dimensions
\cite{dtpin}. In this case, the noise reduction technique helps 
establish the sign of the current on tilted substrate. Without 
noise reduction, the nucleation noise obscures the real sign of the 
current and hence the universality of the model in (2+1) dimensions. 
In particular, for DT model, it has been shown that \cite{dtpin} 
configurations favor downward hops. Thus in spite of intrinsic randomness 
in selecting the neighboring site for a hop, a down-hill current 
is produced on tilted substrate leading to EW type universality.

\subsection{WV Model}
This model was introduced by Wolf and Villain \cite{wv} to simulate  
low temperature MBE growth. In this model, relaxation rules require 
that an adatom will hop to a nearest site if number of $nn$ increases. 
Thus as far as $(0) nn\rightarrow (1)nn$ hops are concerned, the model 
is same as DT model. However, it allows $(1)nn\rightarrow (2)nn$ hops, 
that cause an adatom to dissociate from a step and hop {\it into} the 
surface. Thus clearly, WV model will follow the DT model equation above, in 
addition due to dissociation, downhill current is produced as has 
been discussed in section IV B. The 
current on tilted substrate has been measured for this model and is 
confirmed to be downhill \cite{kps}. 

In (2+1) dimensions, hops from 
lower $nn$ to higher $nn$ imply edge diffusion. This can compensate  
the dissociation induced down hill current. Das Sarma {\it et.al.}
\cite{dtpin} have observed mound formation in (2+1) dimensional WV model.    

\subsection{LD Model}
This model was introduced by Lai-Das Sarma \cite{ld} in connection 
with the LDV equation. The rules were decided, 
based on the geometric interpretation of the term $\nabla^{2}(\nabla h)^{2}$. 
Accordingly, a zero neighbor adatom follows same rules as depicted for DT 
model. If the adatom is deposited at a kink site with a single lateral $nn$, 
it is allowed to move to the nearest kink site with smaller step height. 
Thus, an upward or downward hop is permitted to satisfy the rule. 
The rule suggests that a hop from one kink site to the neighboring 
one is allowed from {\it smaller to larger local slope}. Thus flux
 in expression (1)  
is proportional to $\frac {m}{|m|}\frac {\partial m}{\partial x}$. The 
factor $\frac {m}{|m|}$ ensures proper direction. The probability for 
hopping, once the appropriate configuration is attained is unity 
by relaxation rules for the model. Thus the 
term in the GE due to the movement of adatom at kink is 
$\partial_{x}( \frac {m|m|}{|m|(1+|m|)}
\frac {\partial m}{\partial x})$. The term is consistent 
with the requirement of invariance under $x\rightarrow -x$. For small $m$, 
it reduces to $\frac {\partial^{2}}{\partial x^{2}}(\frac {\partial h}
{\partial x})^{2}$.   
 However this term is expected to contribute mainly 
for larger slopes when the appropriate configurations (steps with 
terraces of unit length appearing consecutively) are large enough in number. 
Under these conditions, the term reduces to, $\partial_{x} (\frac {m}{|m|}
\frac {\partial m}{\partial x})$. This term under scaling 
hypothesis, $x\rightarrow bx$ and $t\rightarrow b^{z}t$ gives exponent $z-3$ 
. If this term is not renormalized, it leads to the 
same scaling exponents as given by  
 $\frac {\partial^{2}}{\partial x^{2}}(\frac {\partial h}{\partial x})^{2}
$ {\it i.e.} $z=3$, $\alpha=1$ and $\beta=1/3$.

\section{Discussion}
Above results show that, a growth situation where kinetics 
of adatoms is well defined can be understood using proposed methodology 
for obtaining the GE. It is therefore perfectly suited 
for computer models with well defined relaxation rules. In real experiments 
as has been pointed out by Krug \cite{kr1} effects of surface roughness, 
evaporation and impurities need to be addressed carefully since these 
can lead to different results for the same system under growth \cite{pk,hm}. 
With this caveat in mind, implications of our results in real MBE like 
growth are discussed.
   
 The results indicate that in real MBE growth, within a low 
temperature range where evaporation is still negligible, one can 
expect different behavior for different materials on a singular 
surface. The processes to 
monitor are in ascending order for activation barrier, SE barrier, 
edge diffusion and dissociation. SE barrier is 
encountered by a single adatom on the surface at the edge while edge 
diffusion is expected to be activated at relatively higher temperature 
due to higher co-ordination around the diffusing adatom. Similarly, 
dissociation is expected to be operative at relatively higher temperatures. 
Thus, for materials with very small or zero SE barrier, at low temperatures 
TI current will dictate the morphology evolution. In many 
cases it will be with $z=4$ and $\beta=1/4$. At higher temperature 
edge diffusion is activated causing instability. This will lead to 
$\beta=1/2$ asymptotically. At higher temperatures, dissociation 
will reduce the uphill current to TI current. However, this 
situation has to compete with the step flow, that leads to EW type growth
\cite{vill1}. 
This scenario is well fitted to growth of Cu \cite{hje}. If the SE barrier 
is high, at low temperature, unstable growth will appear with 
$\beta=1/2$ \cite{hm}. 
At higher temperature, edge diffusion will not change the 
exponent. But once the dissociation is activated, TI current 
will reduce $\beta$. In this study we have neglected any effects 
of upward hops. These can further add more scenarios \cite{svgunpub}. It has 
been mentioned \cite{svgunpub} that with 
in-plane hops $\beta$ cannot exceed the 
value 1/2. The transients in the growth are however known to produce 
higher apparent values of $\beta$ \cite{kr1}.  
Also, upward hops can 
give a value as high as 1 for $\beta$ \cite{svgunpub}. 

\begin{figure} 
\epsfxsize=\hsize \epsfysize= 3.0 in
\centerline{\epsfbox{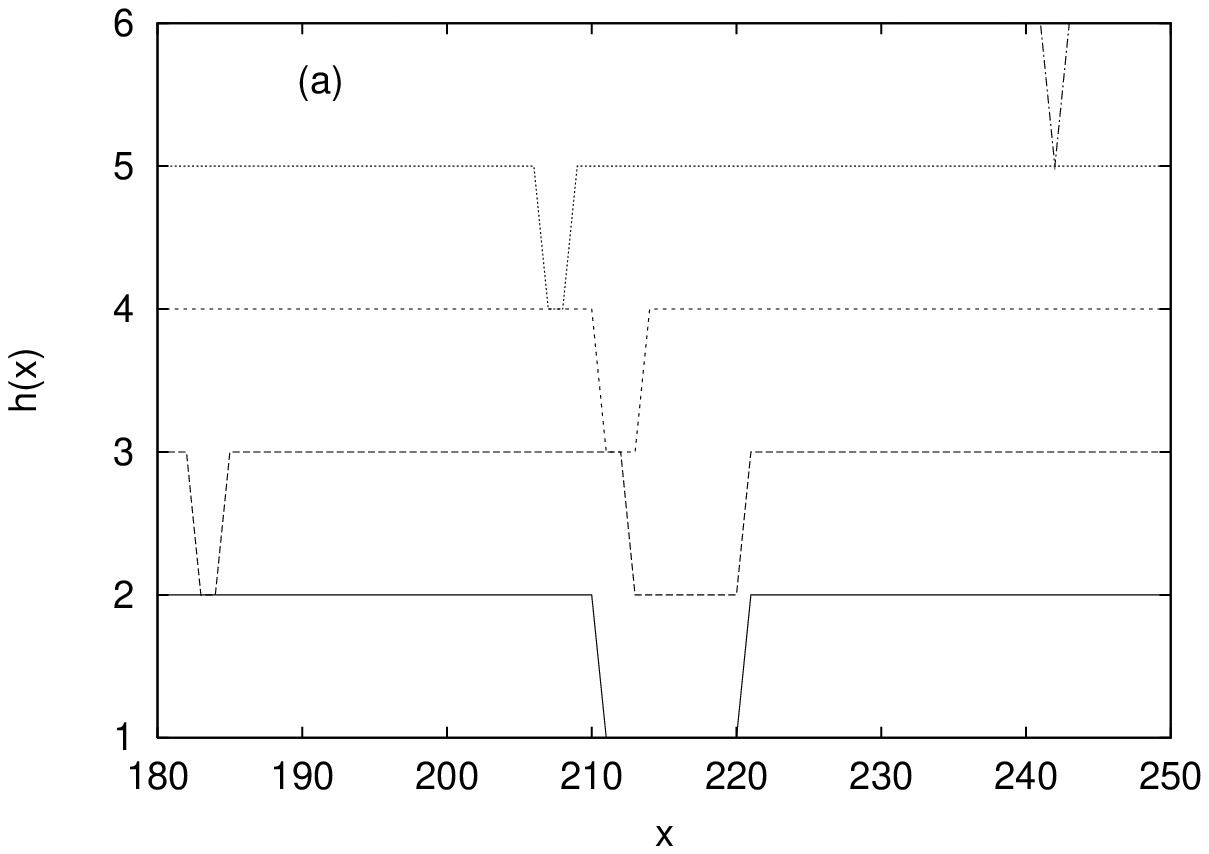}}
\epsfxsize=\hsize \epsfysize = 3.0 in
\centerline{\epsfbox{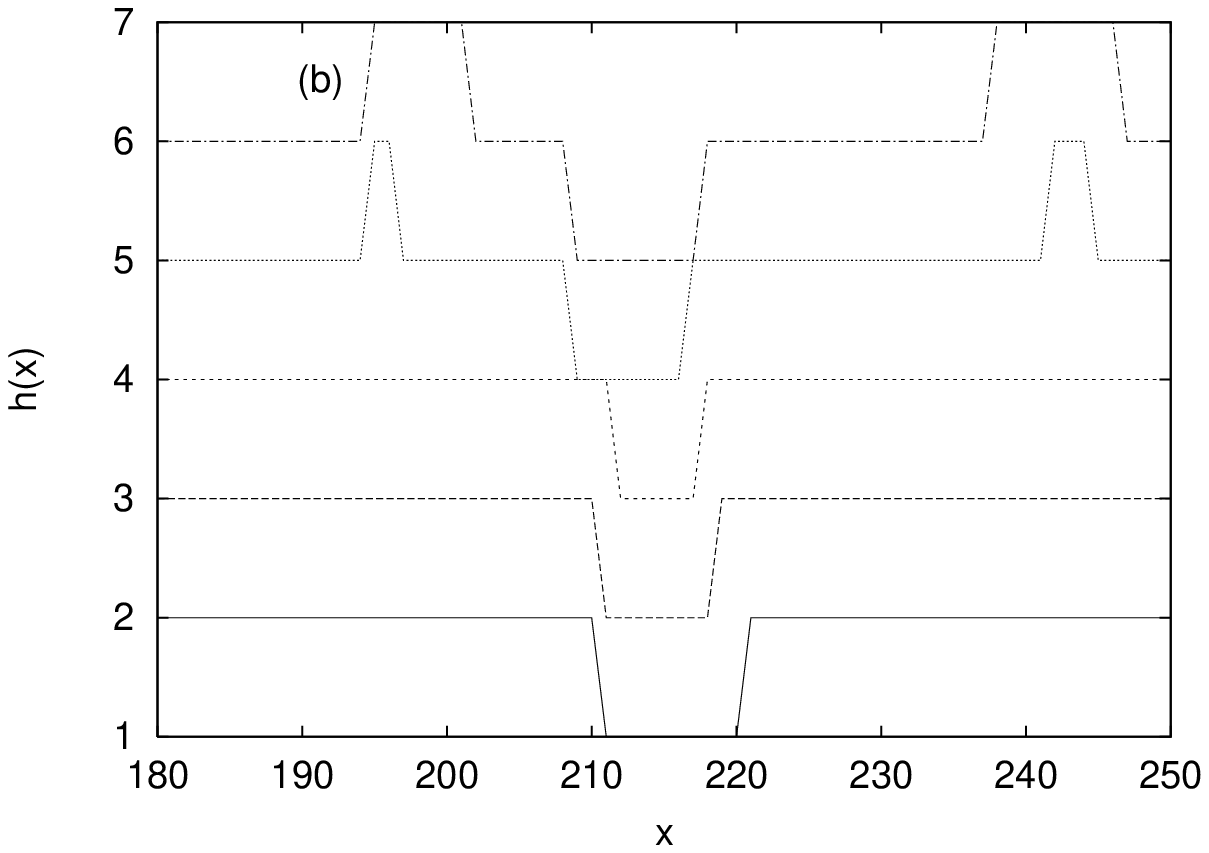}}
\epsfxsize=\hsize \epsfysize = 3.0 in
\centerline{\epsfbox{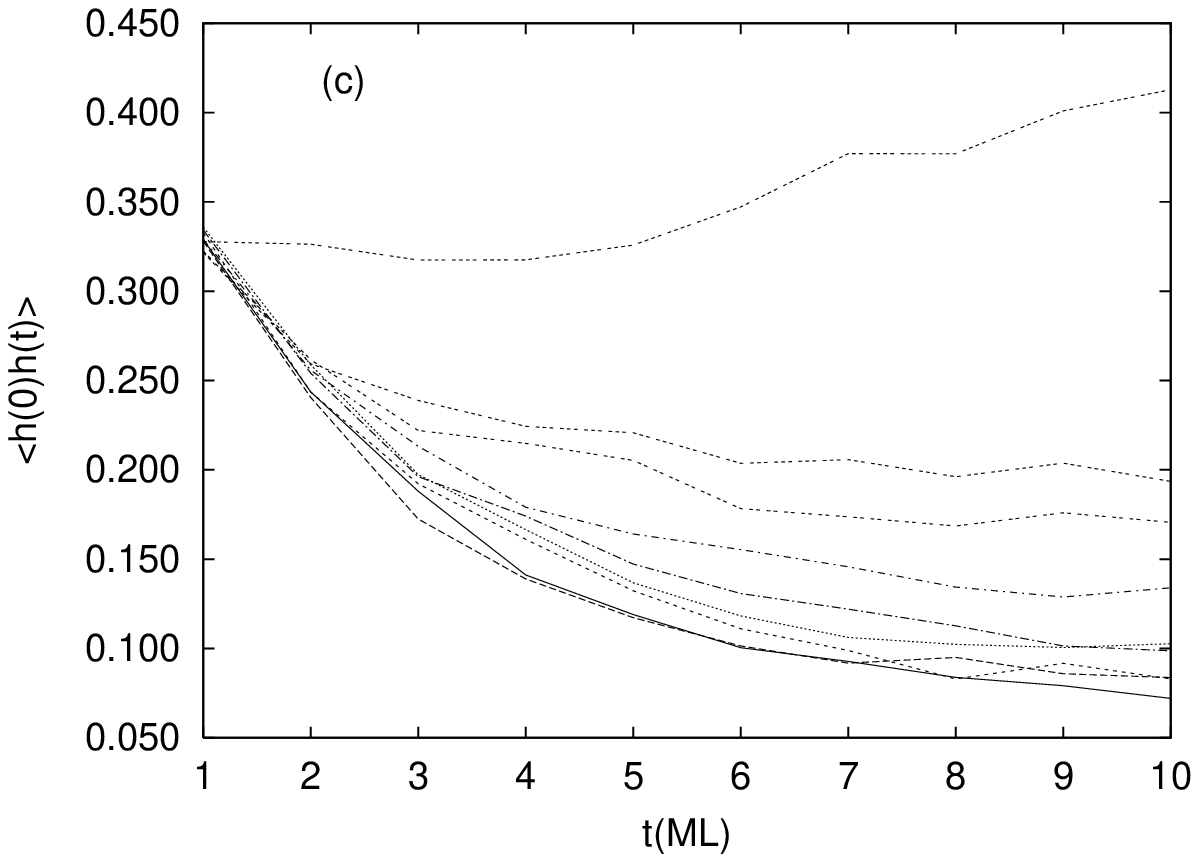}}
\caption{(a) Time development of a base region of width 10 units, 
bounded by single steps of unit height. Figure shows the morphology 
for four layers grown with the parameter $p=0.1$ and $n=100$ for 
the model described in section III.  
 (b) Time development of the base region for four MLs as in (a), 
but the model parameter $p=0.9$ implying large SE barrier. The 
base region is seen to be stable in this case. 
(c) Plot of time correlation function $<h(0)h(t)>$ for the growth 
over base region depicted in (a) and (b). The top most curve correspond to 
the model parameter value $p=0.9$ while curves corresponding to 
$p=$0.8 , 0.7 , 0.6 , 0.5 , 0.4 
, 0.3 , 0.2 , and 0.1 appear below it in the descending order. 
}    
\label{base}  
\end{figure}

So far, we have focused our discussion near the TI current. Above arguments 
suggest that the GE that we have obtained is specific to the 
stepped region on the surface. From the kinetics of adatoms on top or 
base terraces it is clear that the GE will change \cite{svg3}. 
This will lead to the breakdown of spatial invariance. 
We argue that regions that offer restricted types of kinetics 
will support fewer terms in GE than the ones that offer wider possibilities. 
For the sake of argument we will restrict only to (1+1) dimensions, 
however the argument is easily extended in (2+1) dimensions. By inspection, 
a top terrace, defined between two down going step edges, a base region 
defined between two up going step edges, and stepped regions, are three 
distinct regions. Over time scales $<<\tau_{ML}$, a stepped region offers, 
downward hops, relative step motion and in-plane hops. Thus 
equation 6, including all the three terms is valid over this 
region. The top terrace in the absence of nucleation allows downward 
hops. By symmetry, the current must be TI, so that it will support 
only fourth ordered term. On the base region, only in-plane hops are possible. 
Again by symmetry the current must be TI. This region offers Poisson type 
growth with no apparent term to build the h-h correlations. In order that 
such a description is valid on reasonable time scales, it is necessary 
that these regions maintain their identity over appropriate time intervals. 
The corresponding time interval be at least
 $~\tau_{ML}$ which is the 
minimum time for a height fluctuation at a given site. Accordingly, if a 
base region is created locally, then its life time at the given 
place decides whether it will act as an independent region or not. In 
order to get a qualitative idea of such time stability of base regions, 
we have performed simulations of an isolated base region as depicted in 
Fig. \ref {base}(a) and (b). We grow few layers allowing dynamics of 
adatoms as per the growth model described in section III, and 
compute the time correlations 
for height for different values of $p$. 
The width of the base region is of the order of 
diffusion length. The number of hops $n$ are chosen accordingly. We have 
also employed noise reduction technique to reduce the nucleation 
noise. Fig. \ref{base}(a) and (b) show typical development of base 
region for the parameter $p=0.1$ and $0.9$ respectively. Fig.\ref{base}(c)  
shows the time correlations $G_{t}(\tau)=<h(x,0)h(x,\tau)>$, 
for various values of $p$. 
The base region in real growth can occur in various surrounding 
configurations. Although stability times for each such configuration 
will differ, the trend depicted in Fig.\ref{base}(c) is similar with respect 
to the parameter $p$. 
The nature of these curve shows that a) the characteristic time 
$\tau_b$ for the decay depends on $p$ such that, $\tau_b$ increases 
with $p$, and b) for $p=0.9$, $\tau_{b}\rightarrow \infty$.
We find that as long as $p>0.5$ there is a base region with finite depth 
 such that $G_{t}(\tau)$ does not decay. 
For $p=0.9$ this happens for a 
single step depth.  
 Such region offers no kinetics 
that can grow h-h correlations in the vicinity. 
Since no correlations can be built 
in this region, it reduces in size. It acts as a discontinuous region 
with respect to the adjoining stepped regions.  
Corresponding simulations under these conditions 
will always result in the formation of deep ridges.
For such growth the current on tilted substrates is always 
observed to be positive. Above discussion suggests that, when ever, 
$p\neq 0$, initially, there would be regions on the 
substrate during growth separated by  local base regions. 
However, as long as base region decays in time with 
finite $\tau_{b}$ , the lateral growth 
of h-h correlations continues as per the GE on stepped region. 
 In this sense, the GE is valid over the entire 
substrate. However, as growth proceeds, deeper base regions 
will be created by fluctuations. 
If these base regions do not decay in time, which is certain 
when current on tilted substrate is uphill , ridges 
are formed. Lateral growth of the regions separated by a ridge is 
then governed by the dynamics of adatoms across the ridge and 
{\it not by the growth equation on the stepped region.} Thus the 
GE approach fails to describe the growth in such cases.  
 As a result, in (1+1) dimensions 
the mounds grow as $ln(t)$ asymptotically while in (2+1) dimensions 
even slower than $ln(t)$. Power law dependence in time is observed {\it only 
for TI current and downhill current growth} \cite{svg3}.  

In section II we have discussed growth under infinite SE barrier. 
The existence of three regions w.r.t. the GE on a growing interface is 
consistent with the observed growth for infinite SE barrier in (1+1) dimensions 
\cite{kr3}. For the model growth in (1+1) dimensions with the rules in 
section III, unstable growth occurs for $1.0>p>0.5$. Three regions 
are significant w.r.t.the growth under this condition. According to the 
analysis of these regions, the base region supports only Poisson type 
growth while top region can generate $\frac{\partial^{4}h}{\partial x^{4}}$ 
term. Therefore the top regions are flat while base regions have sharp 
ridges breaking $h\rightarrow-h$ symmetry. However, for infinite SE 
barrier, {\it i.e.} for $p=1.0$, top region cannot generate fourth 
ordered smoothening term due to the absence of downward hops. Thus both, 
base and top regions support only Poisson growth rendering symmetric 
pattern.        
\section{Conclusion} 
 In conclusion, we have proposed a simple method for obtaining 
current in a solid-on-solid growth in (1+1) dimension. The resultant growth  
equation shows that presence of diffusion alone is responsible for 
roughening of a singular surface. It induces an asymmetric term in the 
continuum equation. 
The velocity gradient of 
steps on a growing surface is responsible for such a term. In (2+1) 
dimensions, in-plane curvature gradient also generates an asymmetric 
term. This term is independent of SE barrier and is responsible for 
asymmetry in the growth on a two dimensional substrate with infinite 
SE barrier. Role of in-plane hops is seen to smoothen out the deposition 
noise in a plane, within diffusion length. The corresponding terms 
generated are operative only within the plane. 
A curvature 
dependent term is seen to arise from downward hops. Study of zero bias 
model manifests effects of discretization and violation of detailed 
balance. A stable$\rightarrow$ unstable transition with symmetry breaking 
results from such a violation. 
In this connection, present study brings out the effect of dissociation 
on the asymptotic behavior of growth. In the absence of upward hops,
 dissociation introduces a downhill current. The condition of detailed balance 
requires dissociation as a part of the process toward equilibration. Thus 
at high enough temperatures, a zero tilt current is expected to dictate 
the growth morphology. Considering the processes in a KMC simulation it is 
conjectured that these simulations are close to TI current even when SE 
barriers are included in simulations.   
The method is successfully applied to various  models in the literature and 
provides an insight in the model growth. In DT model, in particular it      
supports the dimensional 
dependence of universality class for growth under DT rules.
The GE approach is however seen to be restricted to zero or downhill 
current on tilted substrates. For uphill current, disjoint regions 
following different GEs are obtained.    
 
{\it Acknowledgment:} Author acknowledges useful suggestions by 
Prof. S. Das Sarma, Univ. of Maryland, College Park, U.S.A

\end{document}